\documentclass{nic-series}

\begin{document}

\title{Benchmarking Supercomputers with the J\"ulich Universal Quantum Computer Simulator}

\author{Dennis Willsch \inst{1,2} \and
        Hannes Lagemann \inst{1,2} \and
        Madita Willsch \inst{1,2} \and
        Fengping Jin \inst{1} \and \\
        Hans De Raedt \inst{3} \and
        Kristel Michielsen \inst{1,2}}

\institute{Institute for Advanced Simulation\\
  J\"ulich Supercomputing Centre\\
  Forschungszentrum J\"ulich\\
  52425 J\"ulich, Germany\\
    \email{\{d.willsch, h.lagemann, m.willsch, f.jin, k.michielsen\}@fz-juelich.de}
  \and
  RWTH Aachen University\\
  52056 Aachen, Germany
  \and
  Zernike Institute for Advanced Materials\\
  University of Groningen\\
  Nijenborgh 4, NL-9747 AG Groningen, The Netherlands\\
  \email{deraedthans@gmail.com}
}

\maketitle

\begin{abstracts}
We use a massively parallel simulator of a
universal quantum computer
to benchmark some of the most
powerful supercomputers in the world.
We find
nearly ideal scaling behavior on the Sunway TaihuLight, the K computer,
the IBM BlueGene/Q JUQUEEN, and the Intel Xeon based clusters JURECA and JUWELS.
On the Sunway TaihuLight and the K computer,
universal quantum computers with up to 48 qubits
can be simulated
by means of an
adaptive two-byte encoding to reduce the memory requirements by a factor of eight.
Additionally, we discuss an alternative approach to alleviate the memory bottleneck by
decomposing entangling gates such that low-depth circuits with a much larger number of
qubits can be simulated.
\end{abstracts}

\section{Introduction}\label{willsch_sec_intro}

The simulation of large universal quantum computers on digital computers is a
difficult task since every increase in the number of simulated qubits by one
corresponds to a multiplication of the required amount of memory by a factor of
two. A simulation of a universal quantum computer with 45 qubits requires
slightly more than $0.5$ PB ($0.5\times 10^{15}$ bytes) of memory.  There exist
only a few digital computers with this amount of memory and a powerful network
connecting a large number of compute nodes~\cite{TOP500}. Benchmarking such
systems requires simulation software that can efficiently utilize the
architecture of present day supercomputers. We present benchmarks of some of the
most powerful supercomputers using the J\"ulich universal quantum computer
simulator~(JUQCS). A survey of JUQCS including its instruction set as well as
benchmarks for Shor's algorithm \cite{SHOR99} and the adder circuit
\cite{DRAP00} can be found in Ref.~\citen{JUQCS}. JUQCS has also been used in
the recent quantum supremacy experiments \cite{Google2019QuantumSupremacy}.

In this article, we use the term ``universal quantum computer'' for the
theoretical pen-and-paper version of a gate-based quantum
computer~\cite{NIEL10}, in which the operation of the device is defined in terms
of simple, sparse unitary matrices, representing the quantum gates, acting on
state vectors, without any reference to the real time evolution of physical
systems. In particular, we use the word ``universal'' to refer to a simulation
of the full state vector, independent of the particular quantum circuit,
representing an algorithm in terms of a sequence of gates operated on the
qubits. In addition, we explore a complimentary simulation approach that allows
for an efficient simulation of a much larger number of qubits for low-depth
circuits if only a subset of all amplitudes is required (see also
Refs.~\citen{PEDN17,BOIX17,CHEN18a,CHEN18,MARKOV18,VILLA19,Google2019QuantumSupremacy}).

Since the first massively parallel version of JUQCS was presented in
2007~\cite{RAED07x}, supercomputers have evolved significantly. Therefore, we
have considered it timely to review and improve the computationally critical
parts of the software and use it to benchmark some of the most powerful
supercomputers that are operational today. The characteristics of the
supercomputers that we have used for our benchmarks are summarized in
Table~\ref{willsch_tab_supercomputers}.

\begin{table}[t]
\begin{center}
{\scriptsize
\begin{tabular}{cccccc}
\hline\noalign{\vskip 4pt}
                  & JUQUEEN       & K computer               & Sunway TaihuLight & JURECA & JUWELS \\
\hline\noalign{\vskip 4pt}
 CPU              & IBM PowerPC& eight-core     & SW26010 manycore            &  Intel Xeon    &  Dual Intel Xeon    \\
                  & A2         & SPARC64 VIIIfx & 64-bit RISC                 &  E5-2680 v3    &   Platinum 8168    \\
\hline\noalign{\vskip 4pt}
 clock frequency  &  1.6 GHz      & 2.0 Ghz                  & 1.45 GHz          &  2.5 GHz       &  2.7 GHz    \\
 memory/node      &  16 GB                  & 16 GB          & 32 GB             &  128 GB        &  96 GB   \\
 \# threads/core used  &  1 -- 2                 & 8              & 1                 &  1 -- 2        &  1 -- 2    \\
 \# cores used    &  1 -- 262144            & 2 -- 65536     & 1 -- 131072       &  1 -- 6144     &  1 -- 98304  \\
 \# nodes used    &  1 -- 16384             & 2 -- 65536     & 1 -- 32768        &  1 -- 256      &  1 -- 2048    \\
 \# MPI processes used &  1 -- 524288            & 2 -- 65536     & 1 -- 131072  &  1 -- 1024     &  1 -- 2048    \\
 \# qubits        &  46 (43)                & 48 (45)        & 48 (45)           &  43 (40)       &  46 (43)       \\
\hline
\end{tabular}
}
\end{center}
\caption{Overview of the computer systems used for benchmarking.
The IBM Blue Gene/Q JUQUEEN~\cite{JUQUEEN} (decommissioned),
JURECA~\cite{JURECA}, and JUWELS~\cite{JUWELS} are located at the J\"ulich Supercomputing Centre in Germany,
the K computer at the RIKEN Center for Computational Science in Kobe, Japan,
and the Sunway TaihuLight~\cite{SUNWAY} at the National Supercomputer Center in Wuxi, China.
The row ``\# qubits'' gives the maximum number of qubits that can be simulated with JUQCS--A (JUQCS--E). On JUWELS, the maximum number of qubits was limited to 43 (40)
at the time of running the benchmarks.
}
\label{willsch_tab_supercomputers}
\end{table}

JUQCS is portable software for digital computers and is based on Fortran 2003 code that was
developed in 2007~\cite{RAED07x}. The revised version includes new operations to implement error-correction schemes
and a parser for circuits specified in new quantum assembly flavors such as OpenQASM~\cite{IBMQE,CROS17}.
JUQCS converts a quantum circuit into a form that is suitable for the simulation
of the real-time dynamics of physical qubit models, such as NMR quantum computing~\cite{RAED02}
or quantum computer hardware based on superconducting circuits~\cite{WILL17a}.

The present version of JUQCS comes in two flavors.
The first version, denoted by JUQCS--E, uses double precision (8-byte) floating point arithmetic and can be considered numerically exact (indicated by the E in the acronym).
It has been used to simulate universal quantum computers with up to 45 qubits.
The 45 qubit limit is only set by the amount of RAM memory available on the supercomputers listed in Table~\ref{willsch_tab_supercomputers} (a universal simulation of $N$ qubits using JUQCS--E requires slightly more than $2^{N+4}$ bytes).

A second version, denoted by JUQCS--A, trades memory for additional CPU time and
has been used to simulate a universal quantum computer with up to 48 qubits.
JUQCS--A uses adaptive coding to represent each amplitude of the quantum state
with only two bytes~\cite{JUQCS}. This effectively reduces the memory requirements by a factor of eight relative to the one of JUQCS--E.
The price to pay is a slightly longer execution time and a somewhat reduced numerical precision.
However, we have found that the reduced precision (about 3 digits)
is sufficiently accurate for all standard quantum circuits~\cite{JUQCS}.

We use the acronym JUQCS to refer to both versions of the software, while
JUQCS--E and JUQCS--A are used to specifically refer to the numerically exact version
and the version using adaptive coding, respectively. Since portability was an important design objective, we have so far refrained from using machine-specific programming. A separate version of JUQCS--E utilizing the potential of GPUs is under development.

The memory bottleneck can be alleviated by
decomposing entangling two-qubit gates (entangling gates)~\cite{JUQCS}. This trick
can be used to great advantage if the number of entangling gates is
not too large  and if only a
few of the coefficients of the final state vector need to be computed.
The same idea has proven  to be very useful in quantum Monte Carlo simulations~\cite{HIRS83}. Similar approaches have also been explored by other groups
to simulate large random circuits with low depth~\cite{PEDN17,BOIX17,CHEN18,MARKOV18,VILLA19}.

\section{Simulating Universal Quantum Computers with JUQCS}

In quantum theory, the state of a single qubit is represented by
two complex numbers $a_0$ and $a_1$ which are normalized such that $|a_0|^2+ |a_1|^2 =1$.
A gate operation on the qubit changes these numbers according to
\begin{eqnarray}
\left(
\begin{array}{c}
a_0\\
a_1\\
\end{array}
\right)
\leftarrow
U
\left(
\begin{array}{c}
a_0\\
a_1\\
\end{array}
\right)
,
\label{willsch_eq_gateoperation}
\end{eqnarray}
where $U$ is a unitary $2\times2$ matrix. The state of $N$ qubits is represented by a vector of $2^N$ complex numbers.
Gate operations involving $N$ qubits correspond to matrix-vector multiplications
involving $2^N\times2^N$ unitary matrices (see Ref.~\citen{NIEL10} for more
information).

Typically, an $N$-qubit gate is expressed
in terms of single-qubit gates (i.e., $2\times2$ matrices such as $U$ in
Eq.~(\ref{willsch_eq_gateoperation})) or two-qubit gates, which only act on a subset of the qubits.
Therefore, the matrices required to simulate $N$-qubit circuits are
extremely sparse.
In other words, a quantum gate circuit for a universal quantum computer
is, in essence, a representation of several extremely sparse matrix-vector
operations.
This means that only a few arithmetic operations are required to update each of the
$2^N$ coefficients of the state vector. Therefore, in principle, simulating universal quantum computers is rather simple
as long as there is no need to use distributed memory and the access
to the shared memory is sufficiently fast~\cite{RAED06,SMEL16,KHAM17,HANE17}.
In practice, the speed to perform such operations is mainly limited by the bandwidth to (cache) memory.

For a large number of qubits, however, the only viable way to keep track of the $2^N$ complex coefficients
is to use distributed memory, which comes at the expense of overhead due to communication between nodes,
each of which can have several cores that share the memory (as is the case for all machines listed in Table~\ref{willsch_tab_supercomputers}).
This makes up the ``complicated'' part of JUQCS that implements the MPI communication scheme.
As described in Ref.~\citen{RAED07x}, JUQCS reduces the communication overhead
by minimizing the transfer of data between nodes.
As JUQCS can be configured to use a combination of OpenMP and MPI to
especially tax the processors, the memory, the communication network,
or any combination of these, it provides a practical framework to benchmark high-performance computers.

\section{Validation and Benchmarking}

To validate the operation of JUQCS, we have executed standard quantum algorithms as
well as random circuits using all gates from the JUQCS instruction set for $N\le 30$
qubits on both Windows and Linux systems. Validating the
operation of JUQCS--E (JUQCS--A) when $N$ reaches the limits $N=45$ ($N=48$), set by the amount of
RAM available (see Table~\ref{willsch_tab_supercomputers}), is less trivial because of the requirement to use both MPI and OpenMP on
distributed memory systems. For this reason, we made use of quantum
circuits for which the exact outcome is known.
In this article, we only present results for two particular representatives of
such circuits since they are well suited to the purpose of benchmarking
supercomputers. A more extensive discussion of the algorithms that were used to
validate the operation and study the weak scaling behavior of JUQCS
is given elsewhere~\cite{JUQCS}.

The first circuit that we consider is the circuit that creates a uniform superposition over all
$2^N$ basis states by performing a Hadamard gate on each of the $N$ qubits.
Since the number of states is too large to be verified by sampling,
a practical method to check the result is to compute the single-qubit
expectation values $\langle Q_{\alpha}(i)\rangle =
(1-\sigma_i^\alpha)/2$ where $\sigma_i^\alpha$ for $\alpha=x,y,z$
denotes the Pauli matrix on qubit $i$, the exact values being $0$ for $\alpha=x$ and $1/2$ for $\alpha=y,z$.
In this case, also JUQCS--A can compute the
exact result since the encoding scheme is capable of representing the required amplitudes exactly~\cite{JUQCS}.

\begin{figure}
\begin{center}
\includegraphics[width=0.8\hsize]{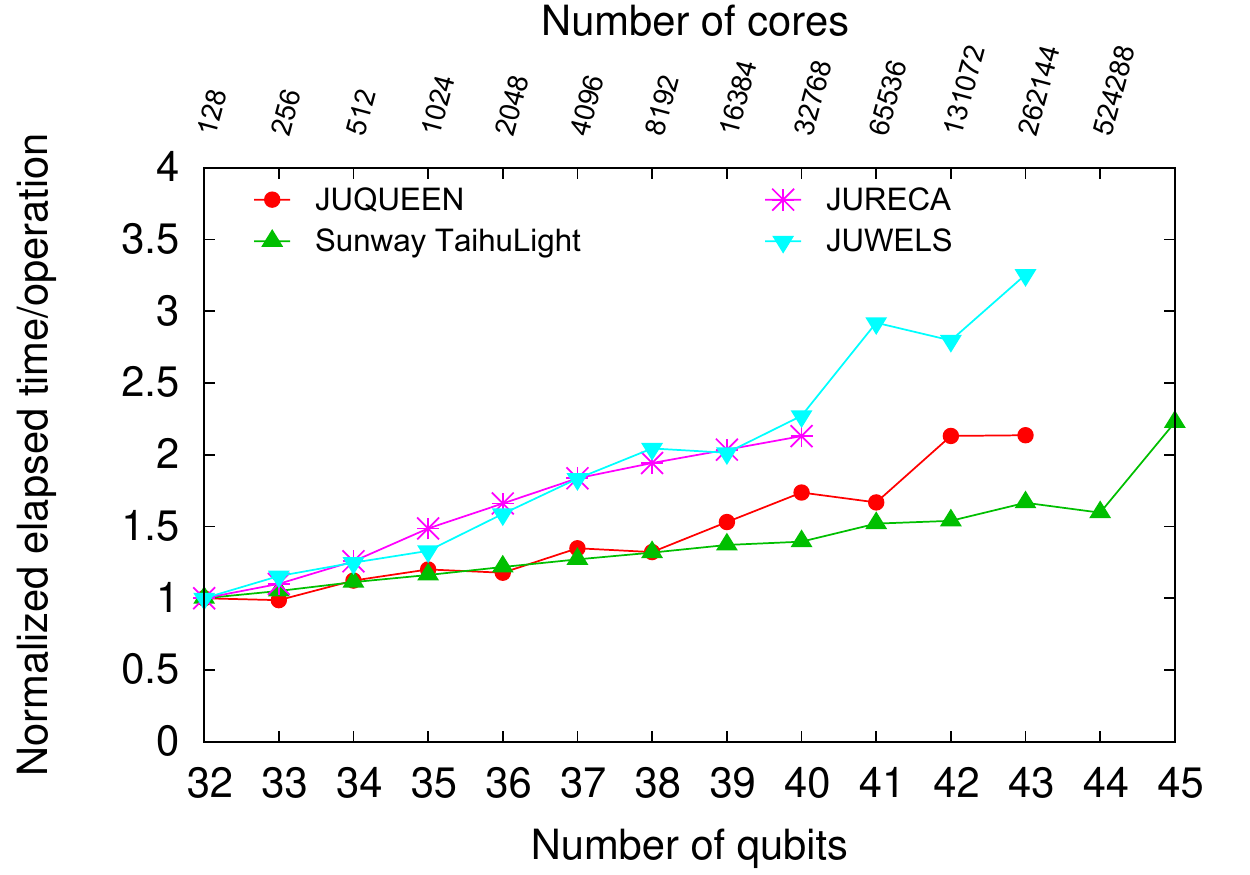}
\caption{
\label{willsch_fig_scaling_JUQCSE}
Weak scaling plot of JUQCS--E executing a Hadamard operation on qubit 0 and the
sequence (CNOT 0 1), (CNOT 1 2), ..., (CNOT N-2, N-1), followed by a measurement
of the expectation values of all qubits.  Shown is the elapsed time per gate
operation as a function of the number of qubits (normalized by the values
corresponding to $N=32$, i.e., $1.0\,\mathrm{s}$ (JUQUEEN), $5.1\,\mathrm{s}$ (Sunway TaihuLight),
$1.4\,\mathrm{s}$ (JURECA), and $0.9\,\mathrm{s}$ (JUWELS)).
}
\end{center}
\end{figure}

\begin{figure}
\begin{center}
\includegraphics[width=0.8\hsize]{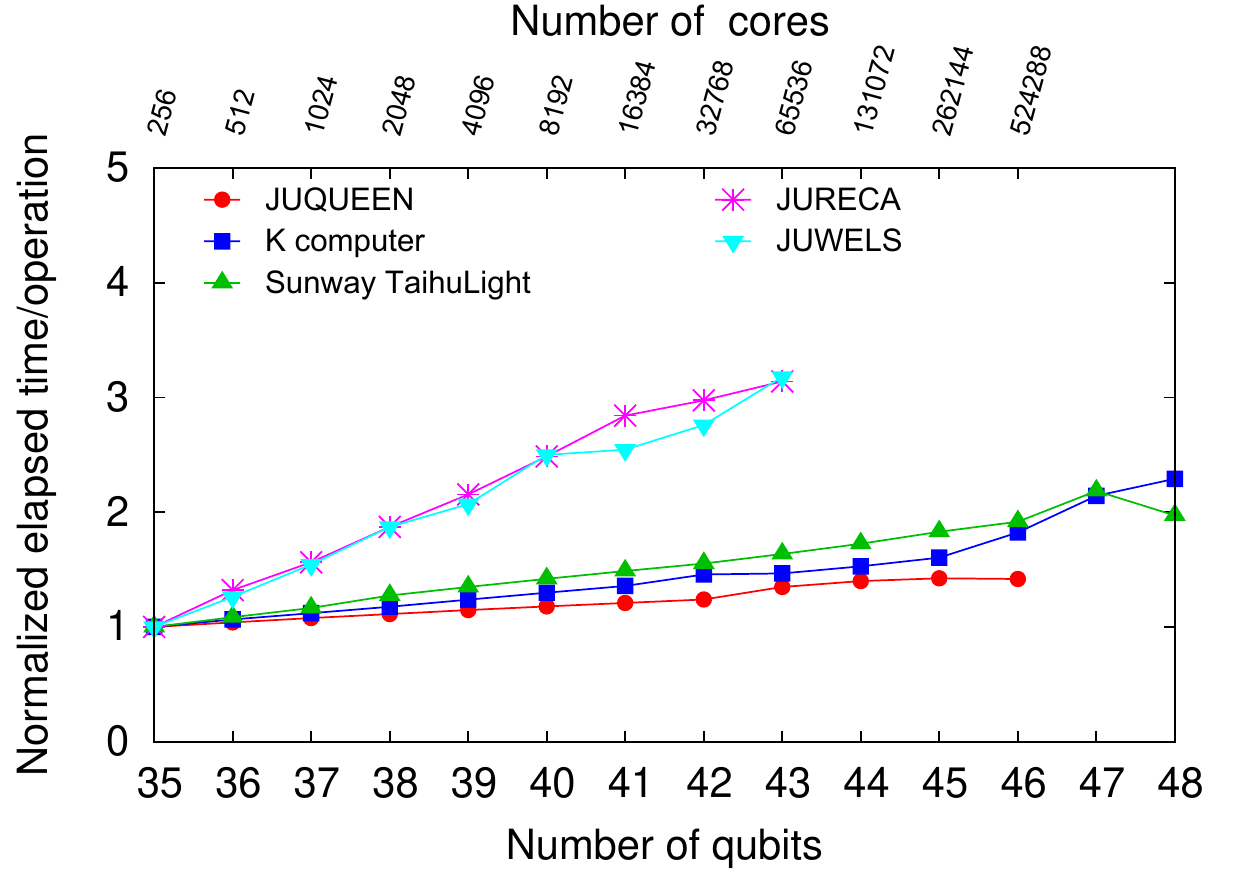}
\caption{
\label{willsch_fig_scaling_JUQCSA}
Weak scaling plot of JUQCS--A executing a Hadamard operation on qubit 0 and the
sequence (CNOT 0 1), (CNOT 1 2), ..., (CNOT N-2, N-1), followed by a measurement
of the expectation values of all qubits.  Shown is the elapsed time per gate
operation as a function of the number of qubits (normalized by the values
corresponding to $N=35$, i.e., $2.7\,\mathrm{s}$ (JUQUEEN), $3.8\,\mathrm{s}$
(K) , $19.9\,\mathrm{s}$ (Sunway TaihuLight), $2.4\,\mathrm{s}$ (JURECA), and
$2.2\,\mathrm{s}$ (JUWELS)).
}
\end{center}
\end{figure}

The second circuit is designed to create the maximally entangled state $(|0\ldots0\rangle+|1\ldots1\rangle)/\sqrt{2}$ by
performing a Hadamard operation on the first qubit and a sequence of
successive CNOT gates on qubits $i$ and $i+1$ for $i=0,1,\ldots,N-2$.
Weak scaling results of executing this circuit on the supercomputer systems
listed in Table~\ref{willsch_tab_supercomputers} are shown in
Fig.~\ref{willsch_fig_scaling_JUQCSE} (JUQCS--E) and Fig.~\ref{willsch_fig_scaling_JUQCSA}
(JUQCS--A).

We see that JUQCS beats the exponential increase in run time with
the number of qubits by doubling the computational resources with each added
qubit.
The weak scaling behavior on the Sunway TaihuLight, the K computer, and JUQUEEN
is close-to-ideal.
However, the weak scaling behavior on JURECA and JUWELS is not as good as the ones on the other supercomputers.
Since the arithmetic work required for the Hadamard gate and the CNOT gate
is rather low, the performance is mainly limited by the memory bandwidth.
This suggests that there may be some limitations in the bandwidth to the memory
and network on JURECA and JUWELS, compared to the other systems used in our benchmark.
The best absolute run time is observed on JUWELS, closely followed by the other
systems except the Sunway TaihuLight, which takes approximately four times
longer.

Comparing
Figs.~\ref{willsch_fig_scaling_JUQCSE} and \ref{willsch_fig_scaling_JUQCSA},
we also find that the computation time for JUQCS--E and JUQCS--A
only differs by a factor of 1--4. The additional time for JUQCS--A
is due to the encoding-decoding operation,
which in turn affects the ratio between computation and communication.
However, the additional time depends on the type of quantum gate. It can range,
e.g.,
from almost zero for the CNOT gate to a factor of 2--3 for the Hadamard gate.
As a result, comparing the computation times of JUQCS--A and JUQCS--E only makes
sense for the same quantum circuit and even then,
because of the difference in the number of qubits and the memory usage,
interpreting differences in the elapsed times is not straightforward.

\section{Memory Reduction by Decomposing Entangling Gates}\label{willsch_sec_entanglement_gates}

There are many ways to alleviate the memory bottleneck. They are typically based
on tensor-network contractions and (Schmidt) decompositions of two-qubit gates
(cf. Refs.~\citen{PEDN17, BOIX17, CHEN18, MARKOV18, VILLA19}). In this section, we adopt an
approach based on the decomposition of entangling gates
using a discrete version of the Hubbard-Stratonovich transformation that
has been used to great advantage in quantum Monte Carlo
simulations~\cite{HIRS83}.

We start by expressing all entangling gates of an arbitrary circuit $C$ in terms
of single-qubit gates and the two-qubit CZ gate, which is always possible~\cite{NIEL10}.
The action of the CZ gate on qubit $i$ and qubit $j$ (denoted by
$\mathrm{CZ}_{i,j}$) is defined as a sign flip of all coefficients with both qubits
$i$ and $j$ in state $\vert1\rangle$. This action
corresponds to the diagonal matrix
$e^{i\pi(1+\sigma_i^z\sigma_j^z-\sigma_i^z-\sigma_j^z)/4}$,
which can be decomposed into a sum of single-qubit operations according to
\begin{equation}
  \label{willsch_eq_decomposition}
  \mathrm{CZ}_{i,j}=\frac{1}{2} \sum_{s \in \{-1,1\}} e^{i(\sigma_{i}^{z}+\sigma_{j}^{z})(x s-\pi/4)},
\end{equation}
where $x$ is a solution of $\cos(2x)=i$.

We partition all qubits $j=0,\ldots,N-1$ into $P$ mutually exclusive subsets
labeled by $p=0,\ldots,P-1$. The dimension of the corresponding subspace is
denoted by $2\le D_p\le 2^N$.
If we decompose all CZ gates between different partitions $p$ according to
Eq.~(\ref{willsch_eq_decomposition}), we can express the circuit $C$ as a sum
of smaller subcircuits,
  $C= \sum_{\mathbf{s} \in \{-1,1\}^{S}} \bigotimes_{p=0}^{P-1} W_{p}(\mathbf{s})$,
where $ S \in \mathbb{N} $ denotes the total number of decomposed CZ gates,
and $W_{p}(\mathbf{s})$ is a subcircuit that only acts on qubits in the partition $p$.
Each subcircuit (a.k.a.~simulation path) can be simulated independently with a quantum computer simulator
such as JUQCS--E or JUQCS--A.
Note that in the extreme case where all CZ gates are decomposed, we have $P=N$,
$D_p=2$, and $W_p(\mathbf s)$ is a product of single-qubit gates on qubit $j=p$ only.

\begin{figure}
\begin{center}
\includegraphics[width=0.8\hsize]{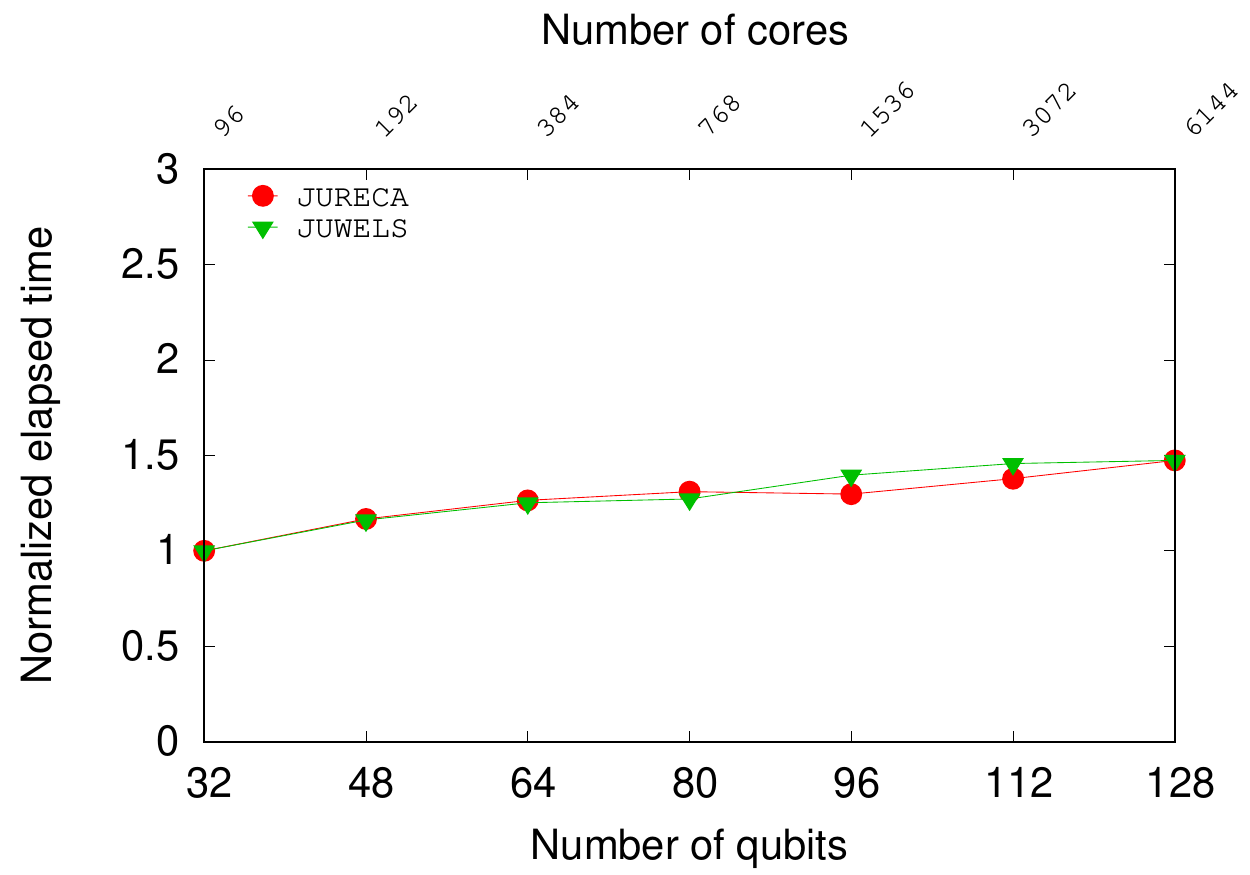}
\caption{
\label{willsch_fig_scaling_lagemann}
Scaling plot of a program implementing the memory reduction scheme
according to Eq.~(\ref{willsch_eq_coeff_expression}), executing
the same circuit as in Figs.~\ref{willsch_fig_scaling_JUQCSE} and
\ref{willsch_fig_scaling_JUQCSA}. Shown is the elapsed time per gate
operation as a function of the number of qubits (normalized by the values
corresponding to $N=32$, i.e., $2.12\,\mathrm{s}$ (JURECA) and $2.10\,\mathrm{s}$ (JUWELS)).
}
\end{center}
\end{figure}

\begin{figure}[ht]
\begin{center}
\includegraphics[width=0.8\hsize]{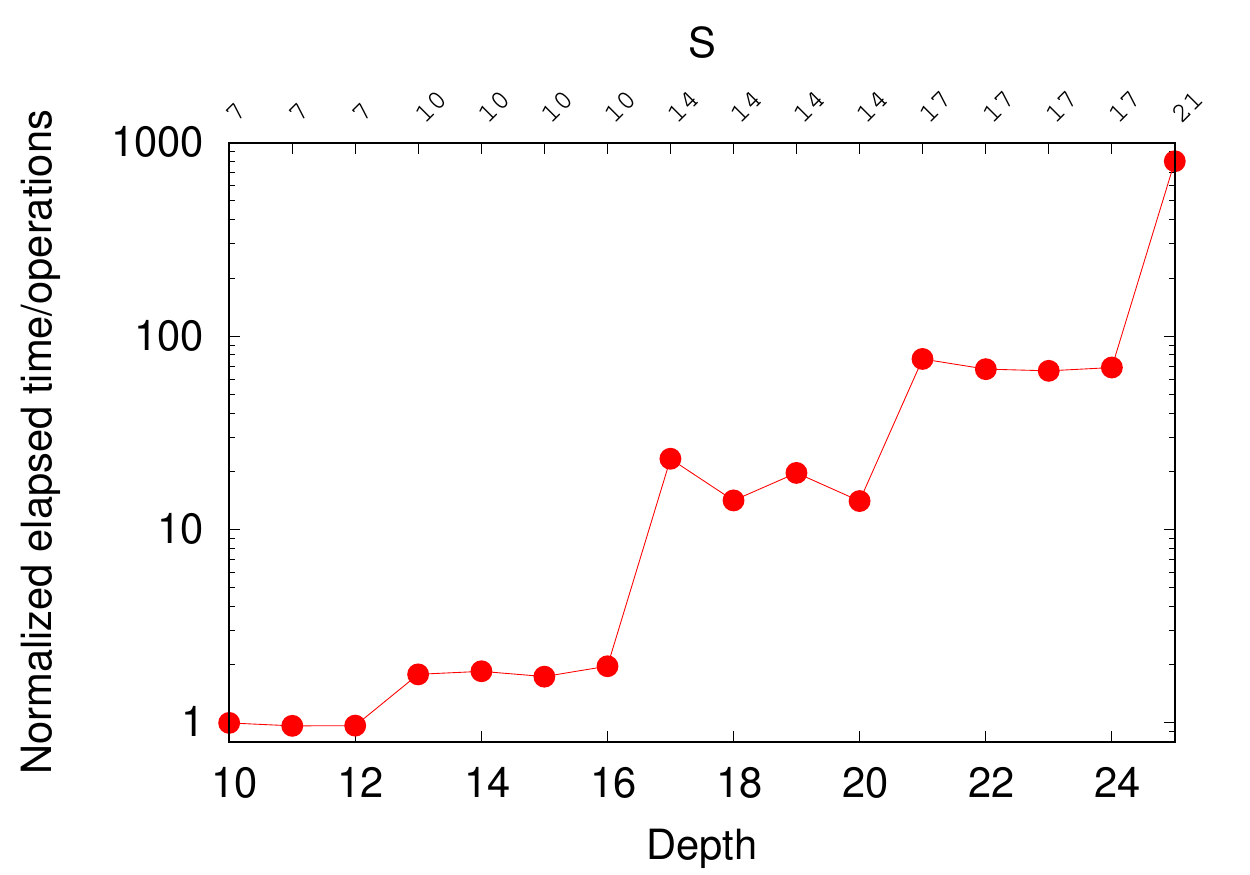}
\caption{
\label{willsch_fig_random_lagemann}
Scaling plot of a program implementing the memory reduction scheme according to Eq.~(\ref{willsch_eq_coeff_expression}), executing a
set of random circuits for 42 qubits as a function of the circuit depth on JURECA. Shown is the elapsed
time per gate operation, normalized by $0.0027\,\mathrm{s}$ corresponding to circuit
depth 10. The axis on top of the figure shows the
number $S$ of decomposed entangling gates.
The total number of gates ranges from 317 (depth 10) to 708 (depth 25). The circuits have been partitioned into two subcircuits of 21 qubits.
The number of extracted coefficients is $M=2^{20}$.
}
\end{center}
\end{figure}

The quantum computer is initialized in the state $\vert\mathbf{0}\rangle$ of the computational basis.
Consequently, the expression for the coefficient corresponding to the bit string $\mathbf{z}$ of the
final state vector is given by
\begin{equation}
  \label{willsch_eq_coeff_expression}
    \langle\mathbf{z}\vert C\vert\mathbf{0}\rangle= \sum_{\mathbf{s} \in \{-1,1\}^{S}}
    \prod_{p=0}^{P-1} \langle\mathbf{z}_{p}\vert W_{p}(\mathbf{s})\vert
    \mathbf{0}_{p}\rangle,
\end{equation}
where $\mathbf{z}_p$ denotes the part of $\mathbf{z}$
belonging to the qubits contained in the partition $p$.

From Eq.~(\ref{willsch_eq_coeff_expression}) it follows that memory reduction can be achieved by separating the circuit into a sum of
$P$ subcircuits acting only on a $D_p$-dimensional subspace of the large $2^N$-dimensional space.
The computation of $\langle\mathbf{z}_{p}\vert W_{p}(\mathbf{s})\vert \mathbf{0}_{p}\rangle$ requires memory bound by the dimension $D_p$ of this subspace.
Obviously, reducing the memory requirements increases the number $S$ of decomposed entangling gates and, consequently, also the computation
time. This increase can be controlled by the choice of the $P$ partitions.

A salient feature of the algorithm is that each term of the sum in
Eq.~(\ref{willsch_eq_coeff_expression}) is independent.  We have parallelized this
sum using MPI to reduce the elapsed time by distributing the work on more cores.
Furthermore, the evaluation of all $W_{p}(\mathbf{s})$'s is
parallelized with OpenMP. If more than one coefficient of the final state vector
is requested, a list of $M$ coefficients $\{\langle\mathbf{z}\vert
C\vert\mathbf{0}\rangle\}$ can be computed in a single run. The reason for this
is that computations for different coefficients only differ by the index of the
coefficient extracted from the vector $W_{p}(\mathbf{s})\vert
\mathbf{0}_{p}\rangle$. Thus $W_{p}(\mathbf{s})\vert
\mathbf{0}_{p}\rangle$ needs to be computed once for each partition $p$ and
$M$ determines the number of coefficients extracted from this result.

In summary, this algorithm has a worst-case time complexity of
$\mathcal{O}(2^{S}P \max\{D_{p},M\}/RT)$ and a worst-case space complexity of
$\mathcal{O}(R\max\{D_{p},M\})$, where $R$ denotes the number of
MPI processes, $T$ denotes the number of OpenMP threads, and $M$ is the desired number
of coefficients from the final state vector.

In Fig.~\ref{willsch_fig_scaling_lagemann}, we present scaling results for the
circuit used for Figs.~\ref{willsch_fig_scaling_JUQCSE} and
\ref{willsch_fig_scaling_JUQCSA}, up to a maximum of $N=128$ qubits
and $M=2^{27}$ extracted coefficients.
The two coefficients corresponding to the states
$\vert\mathbf{0\ldots0}\rangle$ and $\vert\mathbf{1\ldots1}\rangle$ are $1/\sqrt{2}$, as
expected.
Qualitatively, the scaling behavior of the algorithm on JURECA and JUWELS is nearly identical.
Only the data points for the qubit numbers 80, 96, and 112 show small
differences. Since all normalized run times are between 1 and 1.5, both JURECA
and JUWELS show almost ideal scaling, suggesting that the previously observed
limitations do not play a role for this kind of problem.

As the results presented in Fig.~\ref{willsch_fig_scaling_lagemann} are based
on a circuit for which only a small number of entangling gates need to be decomposed, we
also present results for a set of random quantum circuits, generated according
to the procedure given by Boixo et al.~\cite{BOIX18}. The circuits are
characterized by their depth, which is the maximum number of layers (also called
clock cycles or circuit moments) when all consecutive gates on different qubits
are grouped into a single layer. These circuits pose a more difficult problem
for the decomposition algorithm because the control and target qubits of the CZ
gates are distributed in such a way that one cannot partition the circuit into
smaller subcircuits without at least doubling the size $S$ of decomposed
entangling gates.  Furthermore, the density of single-qubit gates is high.
Consequently, at some point, increasing the number of MPI processes will not prevent the elapsed time from growing exponentially. In this respect, all algorithms based on a memory reduction using similar ideas~\cite{PEDN17,BOIX17,CHEN18,MARKOV18,VILLA19} differ from the universal simulator JUQCS in
that the scaling observed for JUQCS is almost independent of the particular circuit simulated.

Figure \ref{willsch_fig_random_lagemann} shows the normalized elapsed run times per operation for random circuit simulations as a function of the circuit depth.
The number of simulated qubits is 42.
We find that for circuits with a low depth, the run time is quasi constant because
enough computational resources are available to distribute the work. For
circuits with depth 13 and larger, we always use
the same amount of computational resources, namely 512 MPI processes and 48
OpenMP threads per process. Consequently, we see a step-like increase in the
run time at depth 17, 21, and 25. These results show that, within reasonable fluctuations, the run time scales
according to the time complexity discussed above when the parameters
$D_{p}$, $R$, and $T$ are held constant and only $S$ changes.

\section{Conclusion}

The massively parallel quantum computer simulator JUQCS has been used to
benchmark the Sunway TaihuLight \cite{SUNWAY}, the K
computer, the IBM BlueGene/Q
JUQUEEN\cite{JUQUEEN}, and the Intel Xeon based clusters JURECA~\cite{JURECA}
and JUWELS~\cite{JUWELS} by simulating quantum circuits with up to $N=48$
qubits. We observed close-to-linear scaling of the elapsed time as a
function of the number of qubits on all tested supercomputers.
The scaling
performance on JUQUEEN, the Sunway TaihuLight, and the K computer tends to be better
than on \mbox{JURECA} and JUWELS, suggesting some limitations in the bandwidth of the latter. The absolute execution times were best on JUWELS, closely followed by JUQUEEN, JURECA, and the K computer. The simulation on the Sunway TaihuLight was approximately four times slower.

Two methods to circumvent the memory problem associated with the simulation of
quantum systems on a digital computer have been explored.
The first uses an adaptive coding scheme to represent the
quantum state in terms of 2-byte instead of 16-byte numbers. We observed that the reduction in memory
has no significant impact on the accuracy of the outcomes (see also Ref.~\citen{JUQCS}).

The second method uses a technique known from Quantum Monte Carlo
simulations to express two-qubit gates in terms of sums of single-qubit gates.
Using this technique, we observed nearly ideal scaling on JURECA and JUWELS for a
maximally entangling circuit with up to $N=128$ qubits. Additionally, we used the method to simulate random 42-qubit circuits up to depth 25.

As the new generation of high-performance computers relies on accelerators or GPUs
to deliver even more FLOPS, we have started to develop a CUDA-based version of JUQCS to explore and benchmark the potential of using GPUs for simulating universal quantum computers.

Since JUQCS can
easily be configured to put a heavy burden on the processors, the memory, the
communication network, or any combination of them, it may be a useful addition to the suite of benchmarks for
high-performance computers.

\section*{Acknowledgments}
We thank Koen De Raedt for his help in improving the JUQCS code.
The authors acknowledge the computing time granted by the JARA-HPC Vergabegremium and provided on
the JARA-HPC Partition part of the supercomputers JURECA~\cite{JURECA} and  JUQUEEN~\cite{JUQUEEN} at the Forschungszentrum J\"ulich.
The authors gratefully acknowledge the Gauss Centre
for Supercomputing e.V. (www.gauss-centre.eu) for
funding this project by providing computing time on the
GCS Supercomputer JUWELS~\cite{JUWELS} at J\"ulich Supercomputing
Centre (JSC).
D.W. is supported by the Initiative and Networking Fund of the Helmholtz Association through the Strategic Future
Field of Research project ``Scalable solid state quantum computing (ZT-0013)''.
Part of the simulations reported in this paper were carried out on the K computer at RIKEN Center for Computational Science in Kobe, Japan,
and the Sunway TaihuLight~\cite{SUNWAY} at the National Supercomputer Center in Wuxi, China.

\bibliographystyle{nic}
\bibliography{database}

\end{document}